\documentclass[a4paper]{article}
\usepackage[a4paper,top=0.75in,bottom=0.75in,left=0.75in,right=0.75in,marginparwidth=1.75cm]{geometry}

\usepackage[english]{babel}
\usepackage[utf8x]{inputenc}
\usepackage[T1]{fontenc}

\usepackage{amsmath}
\usepackage{amssymb}
\usepackage{graphicx}
\usepackage[utf8x]{inputenc}
\usepackage[T1]{fontenc}
\usepackage[colorinlistoftodos]{todonotes}
\usepackage[colorlinks=true, allcolors=blue]{hyperref}
\usepackage{mathtools}
\usepackage{adjustbox}
\usepackage{graphicx}
\usepackage{subfigure}
\usepackage{authblk}
\usepackage{lscape}
\usepackage[outdir=./fig/]{epstopdf}

\def\*#1{\mathbf{#1}}
\def\^#1{\boldsymbol{#1}}

\newcommand{\bmu}{\boldsymbol{\mu}}

\newcommand{\bphi}{\boldsymbol{\phi}}
\newcommand{\bSigma}{\boldsymbol{\Sigma}}

\newcommand{\bLambda}{\boldsymbol{\Lambda}}
\newcommand{\bxi}{\boldsymbol{\xi}}
\newcommand{\bzeta}{\boldsymbol{\zeta}}

\newcommand{\R}{\mathbb{R}}

\newcommand{\EE}[2]{\mathbb{E}_{#1}\left[#2\right]}

\title{A projected nonlinear state-space model for forecasting time series signals}

\author[1,*,$\dagger$]{Christian Donner}
\author[2,*]{Anuj Mishra}
\author[2,3,$\dagger$]{Hideaki Shimazaki}

\affil[1]{Swiss Data Science Center, Zuerich, Switzerland}
\affil[2]{Graduate School of Informatics, Kyoto University, Kyoto, Japan}
\affil[3]{Center for Human Nature, Artificial Intelligence, and Neuroscience (CHAIN), Hokkaido University, Sapporo, Japan}
\affil[*]{These authors contributed equally}
\affil[$\dagger$]{Corresponding authors: chdonner@ethz.ch, h.shimazaki@i.kyoto-u.ac.jp}
\date{}

\begin{document}

\maketitle 

\begin{abstract}
Learning and forecasting stochastic time series is essential in various scientific fields. However, despite the proposals of nonlinear filters and deep-learning methods, it remains challenging to capture nonlinear dynamics from a few noisy samples and predict future trajectories with uncertainty estimates while maintaining computational efficiency. Here, we propose a fast algorithm to learn and forecast nonlinear dynamics from noisy time series data. A key feature of the proposed model is kernel functions applied to projected lines, enabling fast and efficient capture of nonlinearities in the latent dynamics. Through empirical case studies and benchmarking, the model demonstrates its effectiveness in learning and forecasting complex nonlinear dynamics, offering a valuable tool for researchers and practitioners in time series analysis.
\end{abstract}


\section{Introduction}
Learning and forecasting stochastic nonlinear dynamics from noisy observation is imperative in studies involving time series analysis. Examples of time series are the gain returns of stock exchanges or simultaneous activities of multiple neurons to analyze cognitive dynamics. While a wide range of solutions, from the extensions of the classical Kalman filters to modern deep-learning methods, have been proposed, it remains challenging to provide fast and reliable predictions for nonlinear systems when the data is noisy and scarce. 

In this study, we approach this challenge by integrating the low-rank dynamics assumption \cite{thibeault2024low} and suitable regularization for nonlinearity in a state-space modeling framework. We introduce a fast inference and learning algorithm for nonlinear transitions in latent dynamics using univariate Gaussian kernel functions on projected lines in the latent space. Unlike popular kernels such as the Gaussian radial basis functions (RBF) and squared exponential (SE) kernel, we restrict our kernels to be the ridge functions \cite{seshadri2019dimension} to simplify distance calculations in the multidimensional latent space by measuring it along a one-dimensional projection. This way, one can significantly accelerate inference and learning compared to the RBF and SE kernel while potentially capturing the low-rank structure of the dynamics. Our results demonstrate that the proposed model outperforms deep-learning alternatives in predicting chaotic dynamics from a few noisy data samples. Moreover, we showcase that the method more reliably reconstructs the vector fields of the latent nonlinear dynamics than deep-learning methods, indicating superior generalization to the dynamics starting from previously unseen points.

The paper is organized as follows. First, we discuss in section~\ref{sec:related work} related approaches to the problem at hand. In section~\ref{sec:model}, we introduce our proposed state-space model, derive filtering and smoothing equations, and an efficient training algorithm for this model. In section~\ref{sec:results}, we show its forecasting performance on some classical nonlinear systems and its computational efficiency, and then do an exhaustive comparison to a wide range of forecasting methods, from classical to recently peer-reviewed state-of-the-art models, in predicting time series of more than 100 chaotic systems. Furthermore, we compare the methods' capability of reconstructing underlying dynamical systems from noisy data.

\section{Related work}
\label{sec:related work}
In time series modeling, the classical approach for estimating linear systems is the Kalman filter~\cite{kalman1960new,chen2003bayesian}, providing the exact posterior for the linear system with Gaussian noise. For other systems, we generally need to approximate the posterior, either analytically or via sampling. The extended Kalman filter (EKF)~\cite{durbinkoopman2012} and the Unscented Kalman filter (UKF)~\cite{julier2004unscented} are classic examples of the analytic and deterministic sampling-based approaches, respectively. These methods are limited in that they can only provide approximate filter equations when the dynamics' parametric model is known. To overcome the limitation of the parametric models, researchers made considerable efforts to model the dynamics in a nonparametric manner, often using kernel functions or Gaussian processes (GPs). Ghahramani et al. \cite{ghahramani1999} first used kernel functions to model the observation and state models, employed EKF for a linearized model, and further proposed the expectation-maximization (EM) algorithm to learn the model. Ko et al. \cite{ko2007gp, ko2011learning} constructed GP-EKF or GP-UKF that extended EKF or UKF by modeling the observation and transition using GPs. Wang et al. \cite{wang2005gaussian,wang2007gaussian} constructed the GP dynamical model (GP-DM) that finds maximum a posteriori (MAP) estimates of latent states and hyperparameters. To obtain the full posterior distributions, Deisenroth et al. \cite{deisenroth2009analytic} utilized an analytic method called moment matching to construct prediction and filter densities for nonlinear observation and transition modeled by GPs that had the squared exponential kernels as covariance function. For this model, Turner et al.  \cite{turner2010state} employed the moment matching method for inference of filter and smoothing density, and developed an expectation-maximization (EM) algorithm for learning the model parameters. Since then researchers proposed other types of posterior approximation methods such as particle filter \cite{frigola2013bayesian}, expectation propagation \cite{deisenroth2012expectation}, variational Bayes \cite{frigola2014variational}, variational schemes utilizing neural networks \cite{krishnan2015deep, krishnan2017structured, zhao2020variational}. 

Another view of the non-parametric approaches above is to classify them by their form of kernel functions. While the classical approaches used a combination of the RBFs \cite{ghahramani1999, wang2005gaussian, wang2007gaussian, zhao2020variational} to represent the nonlinearity, many of the extended methods used the more general SE kernels \cite{deisenroth2009analytic, 
deisenroth2012expectation, turner2010state, duncker2019learning}. These kernel functions provide a distance metric in the potentially high-dimensional dynamical space. As we will discuss subsequently, both choices suffer from their computational complexity while the SE kernel is more expressive in describing the nonlinearity than the RBF. 

The deep-learning techniques based on artificial neural networks became popular as alternative approaches in time series prediction in recent years. These models include the classical employment of recurrent networks \cite{pearlmutter1989learning,valpola2002unsupervised,honkela2004unsupervised} and the deep-learning techniques (e.g., the long short-term memory (LSTM) \cite{hochreiter1997long,salinas2020deepar}, Transformers \cite{vaswani2017attention,kim2021reversible,liu2023itransformer,d2023odeformer}, and NBEATS \cite{oreshkin2019n}). Recurrent neural networks, best suited for time series or sequence-based data, learn the patterns of sequences by having recurrent connections: the network units are connected to themselves with a time delay. RNNs are popular but difficult to train because the gradients can disappear or explode when learning long-term dependencies \cite{bengio1994learning}. The LSTM model was one of the first techniques that solved the vanishing gradient problem \cite{hochreiter1997long}. Even though new techniques were added, such as better activations, residual layers, batch normalization, dropout, adaptive learning rate schedules, etc., the core architectures stayed the same until the mid-2010s. However, nowadays, the Transformer architecture \cite{vaswani2017attention} is considered to be the preferred model for most sequential modeling tasks. Unlike RNNs that use recurrent connections for information flow over time, transformers use a self-attention mechanism \cite{bahdanau2014NeuralMT} to capture global dependencies across the whole sequence directly. Hence, transformers demonstrate unprecedented performance in sequence-to-sequence tasks and allow for modeling long-term dependencies effectively, avoid vanishing gradient issues common in RNNs, and allow parallel computation during training. Therefore, transformers have become a promising choice for forecasting tasks \cite{kim2021reversible,liu2023itransformer,d2023odeformer}. Separately from these recurrence and attention-based models, a new model NBEATS \cite{oreshkin2019n}, with impressive forecasting ability, that uses a simple but powerful architecture of ensemble feed-forward networks with stacked residual blocks of forecasts and ‘backcasts’ has been introduced recently. 
Unlike probabilistic approaches to modeling the time series, most deep-learning methods do not consider uncertainties in their dynamics and are sometimes limited to forecasting a one-dimensional signal only by default. Nevertheless, we compare our model with these competing architectures and report the competitive forecasting performance of the proposed model.

\section{Model specification}\label{sec:model}
\begin{figure}[t]
\begin{center}
\includegraphics[width=\linewidth]{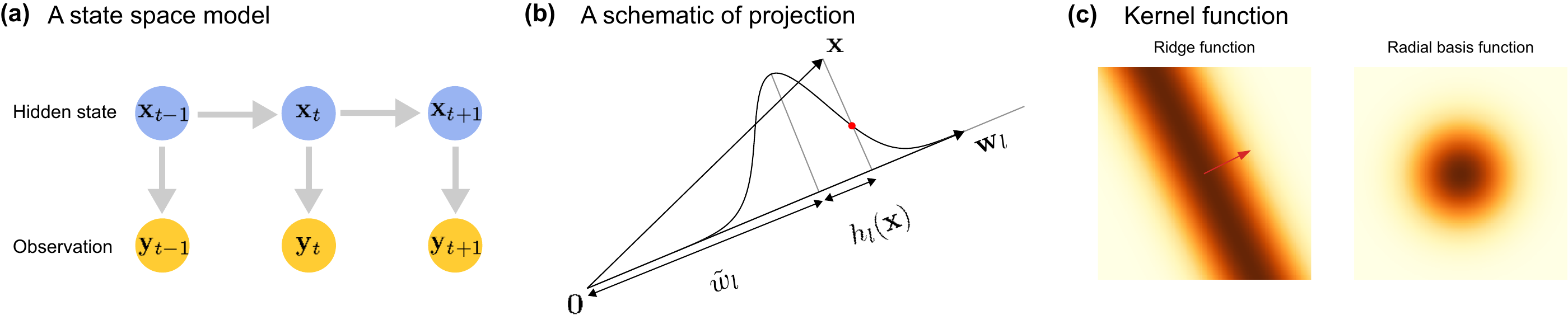}
\caption{A schematic illustration of the projected nonlinear state-space model.
\textbf{(a)} The state-space model is composed of the hidden states and observed variables. 
\textbf{(b)} A schematic illustration of the projection operation, $\phi_{{\rm nl},l}(\*x)$, where the direction of the projection is given by $\*w_l$. The illustration assumes that $\*w_{l}$ is a unit vector.
\textbf{(c)} The ridge function in 2-dimensional latent space (left). For comparison, the RBF kernel is shown on the right. 
}
\end{center}
\label{fig:schamtic_illustration}
\end{figure}

A state-space model is composed of an observation model and a state model (Figure~\ref{fig:schamtic_illustration}\textbf{a}). The former determines the generative process of the observations $\*y_t$ given the latent variables $\*x_t$, while the latter describes the discrete-time dynamics of the latent variable $\*x_t$. Here $\*y_t\in \mathbb{R}^{D_{\*y}}$ and $\*x_t\in \mathbb{R}^{D_{\*x}}$ are the observable and state variables at discrete time $t$ ($t=1,2,\ldots,T$). 

In this study, we assume that the state model is discrete in time, stochastic, and nonlinear:
\begin{equation}
\*x_t = \mathbf{f}(\*x_{t-1}) + \boldsymbol{\xi}_{t},
\label{eq:state_model}
\end{equation}
where $\mathbf{f}: \R^{D_{\*x}} \rightarrow \R^{D_{\*x}}$ is a set of $D_{\*x}$ nonlinear state transition functions, $\*f(\*x)=[f_1(\*x),\ldots,f_{D_{\*x}}(\*x)]^\top$. $\boldsymbol{\xi}_{t}$ represents the latent state noise. We assume zero-mean Gaussian noise, i.e., $\bxi_t\sim {\cal N}(\*0,
\^\Sigma_x)$, where $\^\Sigma_x$ is a $D_{\*x} \times D_{\*x}$ covariance matrix.

We model the nonlinear state transition using kernel functions as follows: 
\begin{equation}\label{eq:state equation}
    \*f(\*x_{t-1}) =
    \begin{pmatrix}
    \*A_{\rm lin} &
    \*A_{\rm nl}
    \end{pmatrix}
    \begin{pmatrix}
    \bphi_{\rm lin}(\*x_{t-1}) \\
    \bphi_{\rm nl}(\*x_{t-1})
    \end{pmatrix}
    + \*b 
    = \*A \bphi(\*x_{t-1}) + \*b. 
\end{equation}
The feature vector $\bphi(\*x)$ is composed of linear features $\bphi_{\rm lin}(\*x) = \*x$ and nonlinear features $\bphi_{\rm nl}(\*x)$, which will be specified using kernels subsequently. Note that, in the case $\*A_{\rm nl}=0$, the standard linear state-space model is recovered. Hence, the nonlinear part extends and compensates for deviations from the linear model. For the nonlinear feature vector $\bphi_{{\rm nl}}(\*x) = [\phi_{{\rm nl},1}(\*x)),...,\phi_{{\rm nl},L}(\*x)]^\top$, we choose a linear projection of the latent variables passed through a squared exponential nonlinearity, i.e., 
\begin{equation}
    \phi_{{\rm nl},l}(\*x) = \exp\left(-\frac{h_{l}(\*x)^2}{2}\right).
    \label{eq:gauss_kernel}
\end{equation}
The input $h_{l}(\*x)$ is given as a projection of the hidden states to the weight vector $\*w_{l}$  (Figure~\ref{fig:schamtic_illustration}\textbf{b}):
\begin{equation}
    h_{l}(\*x) = \*w_{l}^\top \*x - {\tilde w_{l}},
    \label{eq:scaler_input_to_nonlinear_func}
\end{equation}
for $l=1,\ldots,L$. Here $\*w_{l}$ is a $D_{\*x} \times 1$ weight vector and $\tilde w_{l}$ is a scalar constant. Note that $\*w_{l}^\top \*x$ is a projection of $\*x$ to the vector $\*w_{l}$, and $\tilde w_{l}$ forms an offset (inducing) point of the $l$th kernel in this direction (Figure~\ref{fig:schamtic_illustration}\textbf{b}). Each kernel adds $d = D_{\*x} + 1$ parameters to the model. Figure~\ref{fig:schamtic_illustration}\textbf{c} depicts the resulting ridge function in two-dimensional latent space. For a comparison, the RBF is shown on the right panel. Note that the kernel function employed in this study is concentrated only along the projected line and non-concentrated in the orthogonal directions. In contrast, the RBF is concentrated in all dimensions.

The introduced dynamic model can be seen from different perspectives. First, in the light of neural network terminology, it can be interpreted as a residual layer~\cite{he2016deep}. For the non-linear part, the input $\*x_{t-1}$ is transformed by a linear layer, where $\*w_l$ and ${\tilde w_{l}}$ are weights and biases respectively. Then, the linearly transformed input is passed through a non-linearity, i.e., the squared exponential function, and then again linearly weighted by $\*A_{\rm nl}$. The linear part can be seen as a skip connection, which bypasses the non-linearity and is only linearly transformed.

A second viewpoint of the functional choice in Eq.~\eqref{eq:state equation} is in the context of dynamical systems. The dynamics are modeled with a (presumably dominant) linear part, and a nonlinear part models the remaining higher-order terms that correct otherwise a linear model. These higher-order terms appear naturally even in simple systems when the dynamics are modeled in a low-dimensional manifold \cite{thibeault2024low}. Since, in general, the system dimensionality is not known, it is important to account for these non-linearities.

While more complex choices are possible, we assume only a linear observation model in this work:
\begin{equation}
    \*y_t = \*C\*x_t + \*d + \bzeta_t,
\end{equation}
where the noise is again Gaussian, i.e., $\bzeta_t\sim{\cal N}(\*0,\^\Sigma_y)$. Hereafter, we call the proposed model a projected nonlinear state space (PNL-SS) model.

For an observed dataset $\*y_{1:T}$, the model above implies the likelihood
\begin{equation}
    p(\*y_{1:T}\vert \vartheta) = \int d\*x_{0:T} \, p(\*x_0\vert\vartheta)\prod_{t=1}^T  p(\*y_t\vert\*x_t,\vartheta)p(\*x_t\vert\*x_{t-1},\vartheta),
    \label{eq:margina_likelihood}
\end{equation}
where  $\vartheta=\{\*A,\*b,\{\*w_l,\tilde{w}_l\}_{l=1}^L,\^\Sigma_x,\*C,\*d,\^\Sigma_y,\bmu_0,\^\Sigma_0\}$ are the model parameters and $p(\*x_0\vert\vartheta)={\cal N}(\bmu_0, \^\Sigma_0)$ is the initial density.

\subsection{Inference with moment matching} \label{sec:inference}
In this section, we discuss how the posterior over the latent variables $\*x_{0:T}$ can be inferred given data $\*y_{1:T}$ and the model parameters $\vartheta$. As usual for state-space models, we do this first by building the {\it filter density} $p(\*x_t\vert \*y_{1:t},\vartheta)$ for every time-step through forward iterations, and then the {\it smoothing density} $p(\*x_t\vert \*y_{1:T},\vartheta)$  by backward iterations. However, due to the nonlinearity in our model, we cannot solve this exactly, and we resort to an approximation procedure called {\it moment matching}~\cite{deisenroth2009analytic}.

\paragraph{Forward iteration} Let's assume we have the filter density at time-step $t-1$, $p(\*x_{t-1}\vert \*y_{1:t-1}, \vartheta)={\cal N}(\bmu_{t-1}^{\rm f}, \^\Sigma_{t-1}^{\rm f})$ and we wish to compute the density $p(\*x_{t}\vert \*y_{1:t}, \vartheta)$. To do so, we first compute the {\it prediction density}
\begin{equation*}
    p(\*x_t\vert\*y_{1:t-1},\vartheta) = \int p(\*x_t\vert\*x_{t-1},\vartheta)p(\*x_{t-1}\vert \*y_{1:t-1}, \vartheta){\rm d}\*x_{t-1}.
\end{equation*}
Due to the nonlinear mean in Eq.~\eqref{eq:state equation}, however, this integral is not analytically solvable. Luckily, however, we can solve for the mean and covariance of the density above, which are given by
\begin{align}\label{eq:prediction mean}
    \bmu_t^{\rm p} = & \EE{p(\*x_{t-1}\vert \*y_{1:t-1}, \vartheta)}{\*A\bphi(\*x_{t-1}) + \*b}, \\
    \label{eq:prediction covariance}
    \^\Sigma_t^{\rm p} = & \^\Sigma_x + \EE{p(\*x_{t-1}\vert \*y_{1:t-1}, \vartheta)}{(\*A\bphi(\*x_{t-1}) + \*b)(\*A\bphi(\*x_{t-1}) + \*b)^\top} - \bmu_t^{\rm p}(\bmu_t^{\rm p})^\top.
\end{align}
Note that for the expectations above, we need to solve the expectations $\EE{}{\bphi(\*x)}$ and $\EE{}{\bphi(\*x)\bphi(\*x)^\top}$ with respect to a Gaussian density. Because we chose a combination of linear and nonlinear features, where the latter have a Gaussian form with respect to $\*x$, the expectations can be solved analytically (See \ref{appendix:gaussian_integrals}). Using these mean and covariance, we approximate the prediction density with the Gaussian density of the exact first moments, i.e.,
\begin{equation}
    p(\*x_t\vert\*y_{1:t-1},\vartheta) \approx {\cal N}(\bmu_t^{\rm p}, \^\Sigma_t^{\rm p}).
\end{equation}
This approximation is called moment matching. Then, we can compute the (approximate) Gaussian filter density
\begin{equation}
    p(\*x_t\vert\*y_{1:t},\vartheta) = \frac{p(\*y_t\vert\*x_t,\vartheta)p(\*x_t\vert\*y_{1:t-1},\vartheta)}{p(\*y_t\vert\*y_{1:t-1},\vartheta)} \approx {\cal N}(\bmu_t^{\rm f}, \^\Sigma_t^{\rm f}),
\end{equation}
where the approximation is coming from the fact that we plug in the Gaussian approximation of the prediction density (See \ref{appendix:filtering_smoothing}). This completes the forward iteration.

\paragraph{Backward iteration} 

Next, we wish to compute the {\it smoothing densities} $p(\*x_t, \*x_{t+1}\vert \*y_{1:T},\vartheta)$ and $p(\*x_t \vert \*y_{1:T},\vartheta)$. Having $p(\*x_T \vert \*y_{1:T},\vartheta)$ from the last filtering step, we can iterate backward in time and compute
\begin{equation}
    p(\*x_t, \*x_{t+1}\vert \*y_{1:T},\vartheta) = p(\*x_t\vert \*x_{t+1}, \*y_{1:t},\vartheta)p(\*x_{t+1}\vert \*y_{1:T},\vartheta).
    \label{eq:joint_smoothing_posterior}
\end{equation}
The last density is the smoothing density from time $t+1$, and the conditional density is given by
\begin{equation}
    p(\*x_t\vert \*x_{t+1}, \*y_{1:t},\vartheta) = \frac{p(\*x_{t+1}\vert \*x_t,\vartheta)p(\*x_t\vert  \*y_{1:t},\vartheta)}{p(\*x_{t+1}\vert \*y_{1:t},\vartheta)}.
\end{equation}
However, due to the nonlinear mean in Eq.~\eqref{eq:state equation}, this is again intractable, and we perform a Gaussian approximation via moment matching
\begin{equation}
    p(\*x_t, \*x_{t+1}\vert \*y_{1:T}) \approx {\cal N}(\bmu_{t,t+1}^{\rm s}, \^\Sigma_{t,t+1}^{\rm s}),
\end{equation}
where we can calculate $\EE{}{\*x_t}$, $\EE{}{\*x_t\*x_t^\top}$, and $\EE{}{\*x_t\*x_{t+1}^\top}$ analytically with respect to the true density in Eq.~\eqref{eq:joint_smoothing_posterior}. For explicit calculations of moments, see \ref{appendix:filtering_smoothing}.

\subsection{Learning model parameters} \label{sec:learning}
In the previous section, we have seen how to infer an approximate density over the latent variables given the model parameters $\vartheta$. In this section, we utilize the Expectation-Maximization (EM) algorithm~\cite{dempster1977maximum,shumway1982approach} to learn the model parameters. Hence, instead of maximizing the likelihood in Eq.~\eqref{eq:margina_likelihood} directly, we maximize the ${\cal Q}$ function defined as 
\begin{equation}
    {\cal Q}(\vartheta,\vartheta^\prime) \stackrel{\cdot}{=} \EE{\vartheta}{\ln p(\*x_0\vert \vartheta^\prime)} + \sum_{t=1}^T \left( \EE{\vartheta}{\ln p(\*y_t\vert \*x_t, \vartheta^\prime)} + \EE{\vartheta}{\ln p(\*x_t\vert \*x_{t-1}, \vartheta^\prime)} \right),
    \label{eq:qfunction}
\end{equation}
where $\EE{\vartheta}{\cdot}$ refers to the expectation by the smoothing densities. Since the observation model is linear with Gaussian noise, all its parameters $\*C,\*d,\^\Sigma_y$ can be solved analytically. The same holds for the mean and covariance of the initial density $\^\mu_0$ or  $\^\Sigma_0$. Though the state model is nonlinear, many parameters can be optimized analytically as well, thanks to our choice of nonlinearity. The optimal transition matrix and offset are given by (See \ref{appendix:q_function})
\begin{align}
	\*A^* = & \left(\sum_{t=1}^T \EE{}{(\*x_t - \*b)\boldsymbol{\phi}(\*x_{t-1})^\top }\right)\left(\sum_{t=1}^T\EE{}{ \boldsymbol{\phi}(\*x_{t-1}) \boldsymbol{\phi}(\*x_{t-1})^\top}\right)^{-1}, \\
	\*b^* = & \frac{1}{T}\sum_{t=1}^T\EE{}{\*x_t - \*A\boldsymbol{\phi}(\*x_{t-1})}.
\end{align}
Also, we can solve for the optimal covariance matrix
\begin{equation}
	\^\Sigma_x^* = \frac{1}{T}\sum_{t=1}^{T}\EE{}{(\*x_t - \*f(\*x_{t-1}))(\*x_t - \*f(\*x_{t-1}))^\top}.
\end{equation}
The parameters of the feature vector $\^\phi(\*x)$, namely $\{\*w_i,\tilde{w}_i\}_{i=1}^K$, are the only ones that we cannot optimize analytically. We solve them by gradient ascent, maximizing the ${\cal Q}$ function.

\subsection{Computational considerations} Now lastly, we provide a more detailed overview of the computational advantages due to the choice of the nonlinearity by Eqs~\eqref{eq:gauss_kernel} and \eqref{eq:scaler_input_to_nonlinear_func}. For notational convenience, we drop the time index in this subsection. For the moment matching in Sec~\ref{sec:inference} and the computation of the ${\cal Q}$ function in Sec~\ref{sec:learning}, we need to compute integrals of the form
\begin{equation}
	\int g(\*x)\phi_{{\rm nl},l}(\*x)p(\*x){\rm d} \*x = \int g(\*x)\exp\left(-\frac{(\*w_l^\top\*x-\tilde{w}_l)^2}{2}\right)p(\*x){\rm d} \*x,
\end{equation}
where $p(\*x)={\cal N}(\bmu,\^\Sigma)$ and $g(\*x)\in\{1,\*x,\*x\*x^\top\}$. We note that the feature $\phi_{{\rm nl},l}(\*x)$ is proportional to a Gaussian, and hence
\begin{equation}
	\exp\left(-\frac{(\*w_l^\top\*x-\tilde{w}_l)^2}{2}\right)p(\*x)\propto {\cal N}(\bmu_\phi, \^\Sigma_{\phi}),
\end{equation}
where
\begin{align}
	\^\Sigma_{\phi}^{-1} = &\^\Sigma^{-1} + \*w_l\*w_l^\top, \\
	\^\mu_{\phi} = &\^\Sigma_{\phi}\left[\tilde{w}_l\*w_l + \^\Sigma^{-1}\^\mu\right].
\end{align}
We see that the integral is computed with respect to a new Gaussian measure, whose precision matrix $\^\Sigma_\phi^{-1}$ is a rank one update of the matrix $\^\Sigma^{-1}$. Thanks to the Sherman-Morrison formula and the matrix determinant lemma, we can compute the matrix inversion and its determinant efficiently. Both quantities are needed for solving the integral. Naturally, this also extends to integrals, where we need to integrate with respect to a Gaussian measure of the form $\phi_{{\rm nl},l}(\*x)\phi_{{\rm nl},m}(\*x)p(\*x)$, just that we need to perform twice a rank one update or once a rank two update.

This computational advantage is lost with the choice of a Gaussian RBF
\begin{equation}
	\phi_{{\rm nl},l}(\*x) = \exp\left(-\frac{||\*x - \*c_l||^2_2}{2 s_{l}^2}\right),
\end{equation}
where $\Vert \cdot \Vert_2$ is the Euclidian norm. $\*c_l$ and $s_l$ are a location and scale paramter. With this choice, the integrals discussed above are still tractable. However, the computational convenience of the rank one updates is gone as soon as $D_x>1$. This implies that for each integral computation, we need to perform a full matrix inversion. Since we need to compute these integrals numerous times, this considerably slows inference and parameter optimization. Moreover, this effect increases with larger latent dimensions $D_x$. However, note that when $D_x=1$ the models are equivalent, i.e., represent the same model class and have the same computational complexity. 

\section{Performance evaluation}\label{sec:results}

\subsection{Forecasting a nonlinear oscillator}

\begin{figure}[tp]
\centering
\includegraphics[width=\linewidth]{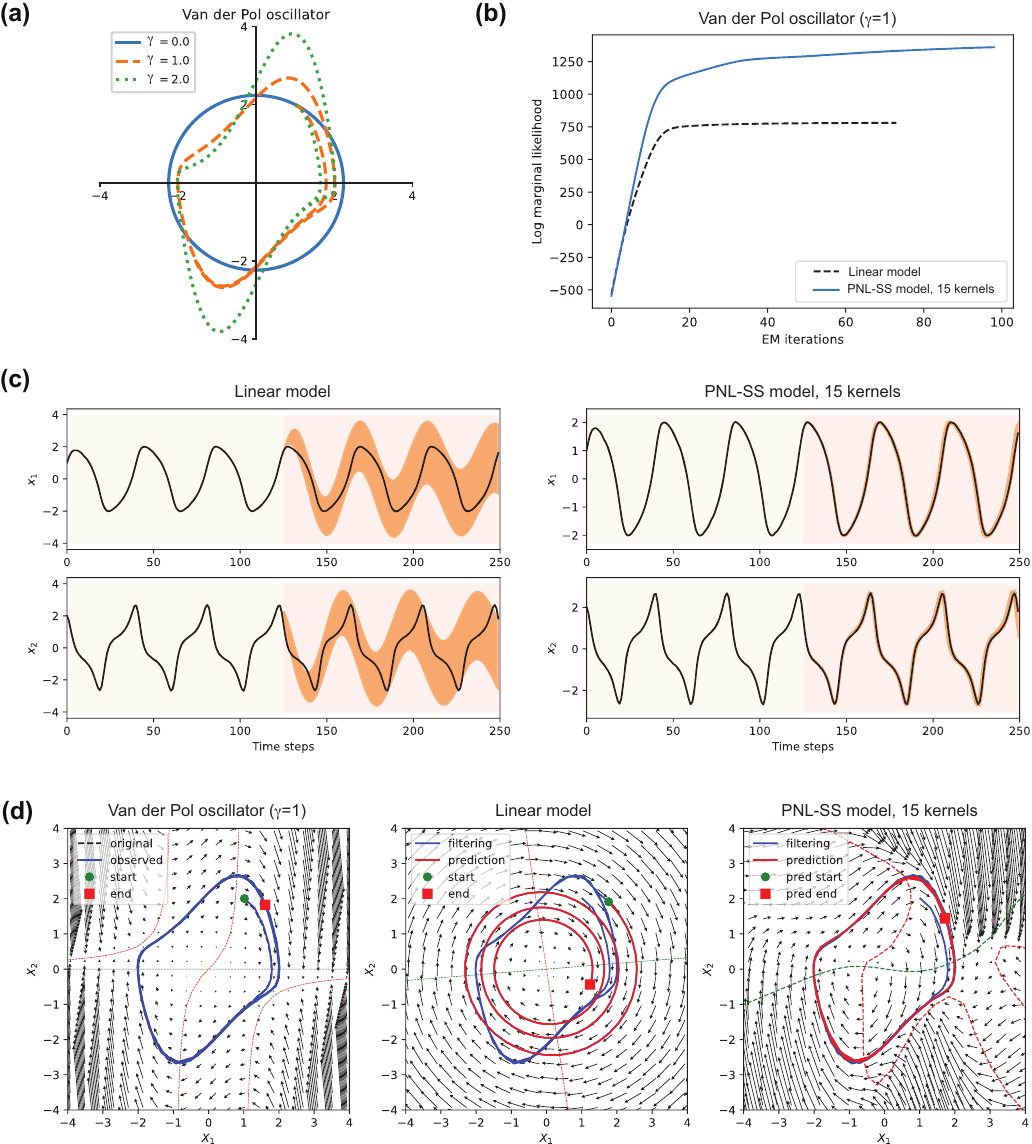}
\caption{Forecasting a Van der Pol oscillator by the linear and projected nonlinear state-space (PNL-SS) models.
\textbf{(a)} The ground truth dynamics of the Van der Pol oscillator with $\gamma=0, 1, 2$, starting at $\*x=(1,2)$. 
\textbf{(b)} Marginal likelihood of the linear (dashed) and projected nonlinear (solid) state-space models as a function of the EM iteration steps. The learning was completed when the increase of the marginal likelihood ratio from the previous value is less than $10^{-4}$ or at the maximum 100 iterations. 
\textbf{(c)} Estimation and prediction of two-dimensional latent states of the Van der Pol oscillator by the linear (left) and projected nonlinear models (right). The data points on time $t=1, \ldots, 125$ are leaned by the models (yellow background). Smoothed estimates of the latent states are given during this period. The prediction was made for $t=126, \ldots, 250$ (light orange background). The orange areas indicate a $95\%$ credible interval of the prediction density. Note that the colored areas are too narrow to be visible during the learning period. 
\textbf{(d)} (left) The vector field (arrows) of the nonlinear oscillator in the two-dimensional space with the original (dashed) and observed dynamics with noise (solid blue). The dashed black lines are behind the blue lines. The dotted red and green lines are nullclines, where the vector field possesses a zero component. 
(middle) The vector field and nullclines are learned by the linear state-space model. The blue and red solid lines are estimated and forecasted trajectories, respectively. 
(right) The same as in the middle panel for the PNL-SS model. 
}
\label{fig:VanderPol}
\end{figure}

First, we apply the PNL-SS model to a noisy Van der Pol oscillator as a prototypical example of a nonlinear oscillator. The Van der Pol oscillator without noise is governed by
\begin{align}
    \dot x_1 &= x_2, \label{eq:VanderPol1}\\
    \dot x_2 &= \gamma (1 - x_1^2) x_2 - x_1.
    \label{eq:VanderPol2}
\end{align}
We performed numerical integration of the nonlinear differential equations without adding noise to the dynamical equations. The dynamics of the nonlinear oscillator for different values of $\gamma$ are shown in Figure~\ref{fig:VanderPol}\textbf{a}. The observations of the dynamics were made at equidistant $T=250$ points in the interval [0, 40]. For simplicity, we use discrete points to denote the time steps, i.e., $t=1,\ldots,T$. The independent Gaussian noise with a small variance of $\sigma_{\rm obs}^2=0.0001$ ($\^\Sigma_y = \sigma_{\rm obs}^2 \*I$) was added to the generated time series to obtain the observed time series.

We fitted the linear and PNL-SS models with the latent dimension $D_x = 2$ to the $D_y=2$-dimensional observation. The example shown here used $L=15$ kernels. To simplify the interpretation of the latent processes, we fixed $\*C=\*I$ and $\*d=\*0$ in this analysis. For comparison, we also fitted a linear state-space model (Kalman filter). The marginal likelihood of the PNL-SS model exceeds that of the linear model as expected (Figure~\ref{fig:VanderPol}\textbf{b}), and the difference was significant (chi-squared test, $p<0.01$). In Figure~\ref{fig:VanderPol}\textbf{c}, we used the learned models to estimate the underlying dynamics (left, linear model; right PNL-SS model) from the noisy observation for $t=1,\ldots,125$, and forecasted the dynamics for $t=126,\ldots,250$. The solid lines are MAP estimates, and the orange area indicates $95\%$ credible intervals, which shows the superior performance of capturing nonlinear dynamics by the PNL-SS model. Another goal of learning the nonlinear dynamics is to analyze its vector field and nullclines to understand emergent nonlinear phenomena \cite{ishii2001reconstruction}. Figure \ref{fig:VanderPol}\textbf{d} shows the original and learned vector fields with the forecasted trajectories. While the linear model can construct the circular vector field only, the reconstructed vector fields and nullclines of the PNL-SS model faithfully capture the original ones, confirming that the learned model can forecast trajectories starting from previously unseen initial points.

\subsection{Comparison with a state model using RBF kernel}

The standard approach for modeling nonlinear dynamics using the basis functions or Gaussian processes uses the RBFs or their extensions. Both the projected kernel and RBF have the same number of parameters. The projection in each kernel  Eq.~\eqref{eq:scaler_input_to_nonlinear_func} contains $D_x$ parameters for $\*w_l$ and a scalar inducing point $\tilde w_l$. The RBF kernel has $D_x$ parameters for $\*c_l$ and a scalar parameter $s_l$. However, as discussed earlier, the rank-one structure in the kernels of the PNL-SS model offers advantages in computational cost in forward inference and learning, thanks to the Sherman-Morrison formula. 

\begin{figure}[tp]
\centering
\includegraphics[width=\linewidth]{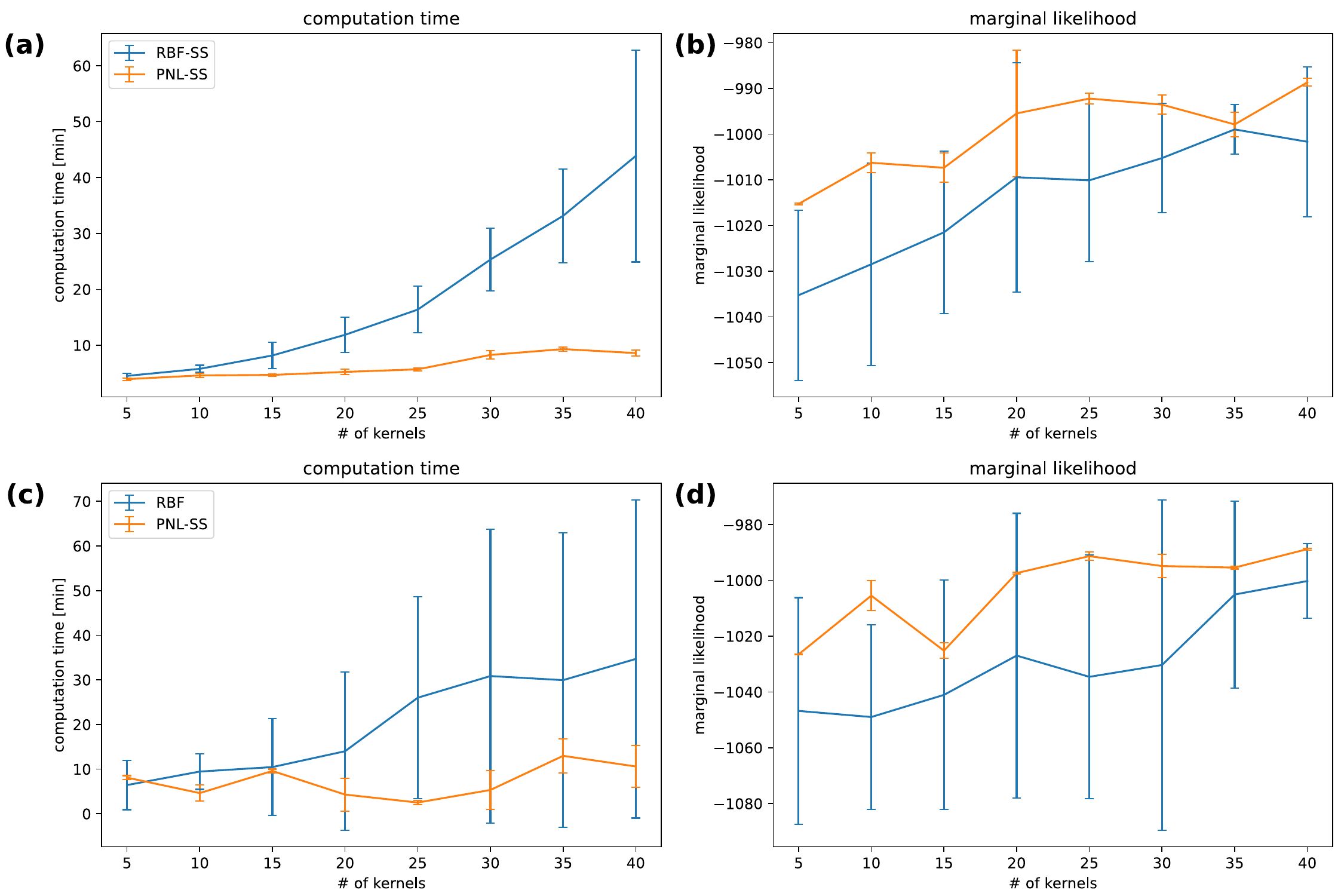}
\caption{Comparison of the PNL-SS and RBF-SS models in fitting a noisy Lorentz attractor.
\textbf{(a)} Computational time required for completing 50 EM iterations for the PNL-SS (orange) and RBF-SS (blue) as a function of the number of kernels. The line graph for each method shows the mean for fitting 10 sample trajectories with 2SD error bars. Computation was performed by Intel Xeon 6258R 2.7G with Tesla V100s using a custom code powered by Jax.
\textbf{(b)}
Marginal likelihoods of the PNL-SS (orange) and RBF-SS (blue) models after completing 50 EM iterations as a function of the number of kernels. Note that the PNL-SS and RBF-SS models possess an equal number of parameters. 
\textbf{(c)} Computational time for each method until the same convergence criteria are satisfied. The convergence is achieved if the increase of the marginal likelihood ratio from the previous value is less than $10^{-4}$ or reaches the maximum EM iteration of 100. 
\textbf{(d)} Log marginal likelihoods of each method after the convergence or maximum 100 iterations.
}
\label{fig:lorentz_nlss_rbf}
\end{figure}

Here, we investigated the computational time and goodness-of-fit of these two models by fitting them to the noisy Lorentz system, i.e., 3-dimensional chaotic attractor dynamics, while varying the number of kernels used in the models. We fitted the PNL-SS and RBF-based model to 300 time points of Lorentz dynamics with additive independent Gaussian noise of small variance ($\sigma_{\rm obs}^2=0.0001$). The PNL-SS method is significantly faster than the RBF-based method, assessed by the computational time required to either complete 50 EM iterations or approach the convergence criteria (Figure~\ref{fig:lorentz_nlss_rbf}\textbf{a,c}). Furthermore, while having the same number of parameters, the PNL-SS model outperformed the RBF-based method in terms of the marginal likelihood regardless of the number of kernels (Figure~\ref{fig:lorentz_nlss_rbf}\textbf{b,d}).

\subsection{Benchmarking against other forecasting methods} 

\begin{figure}[tp]
\centering
\includegraphics[width=\linewidth]{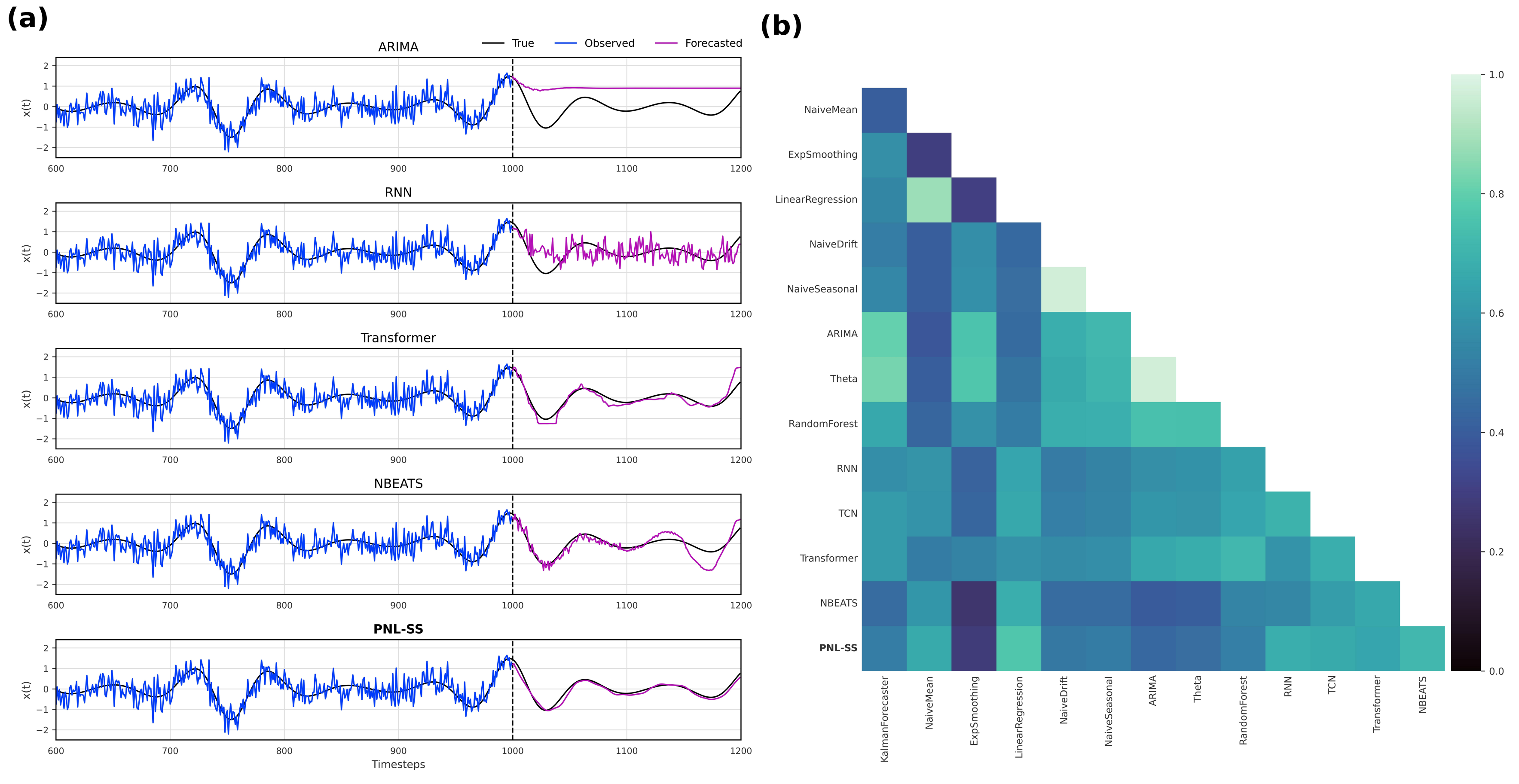}
\caption{Comparison of methods in forecasting noisy, chaotic time series. \textbf{(a)} Forecasting results of AIZAWA dataset using ARIMA, LSTM-RNN, Transformer, NBEATS, and PNL-SS methods. The models fitted to noisy observation of AIZAWA (original: smooth black lines, noisy observation: blue lines) for the first 1000 time points were used to forecast the following 200 time points (purple lines). The additive observation noise with the variance $\sigma_{\rm obs}^2=0.64$ was added to the original dynamics. 
\textbf{(b)}
Spearman correlation of forecasting errors among different forecasting methods computed across all dynamical systems.
}
\label{fig:combine1}
\end{figure}

\begin{figure}[th]
\centering
\includegraphics[width=\linewidth]{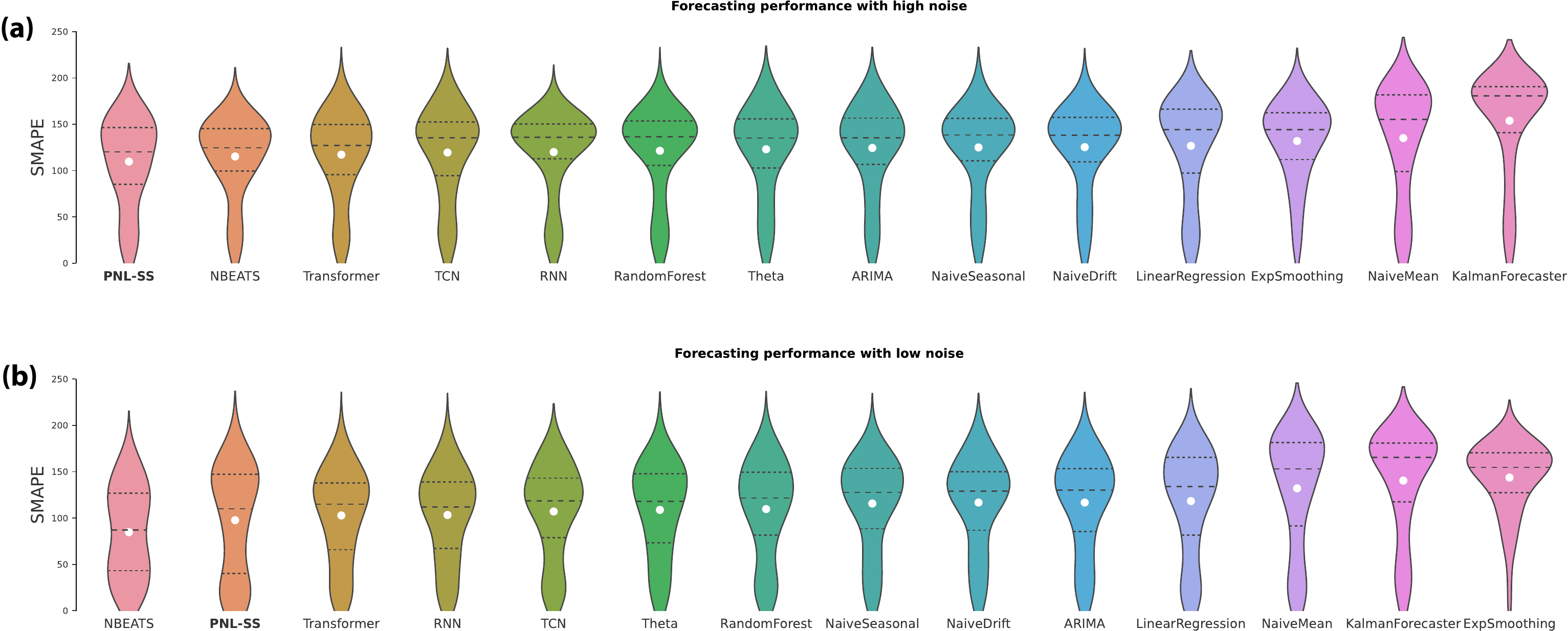}
\caption{Forecasting results with high and low noise for chaotic time series data. 
\textbf{(a)} The distributions of forecast errors for 126 dynamical systems with high observation noise across different forecasting models, sorted by increasing mean error. Independent Gaussian noises with the zero mean and variance $\sigma_{\rm obs}^2 = 0.64$ were added as the observation noise in the high noise condition.  Here, the forecast error was measured by the symmetric mean absolute percentage error (SMAPE). See~\ref{appendix:forecasting details} for details. The white dot within each violin plot shows the mean of the SMAPEs. The dotted horizontal lines show quartiles, where the middle one is the median.
\textbf{(b)} The forecasting results for low observation noise ($\sigma_{\rm obs}^2 = 0.04$). Note that the model order along the horizontal axis differs between the two panels because the relative performance of different forecasting methods changes with levels of noise present.
}
\label{fig:violin_high_low}
\end{figure}

\begin{figure}[tp]
\centering
\includegraphics[width=\linewidth]{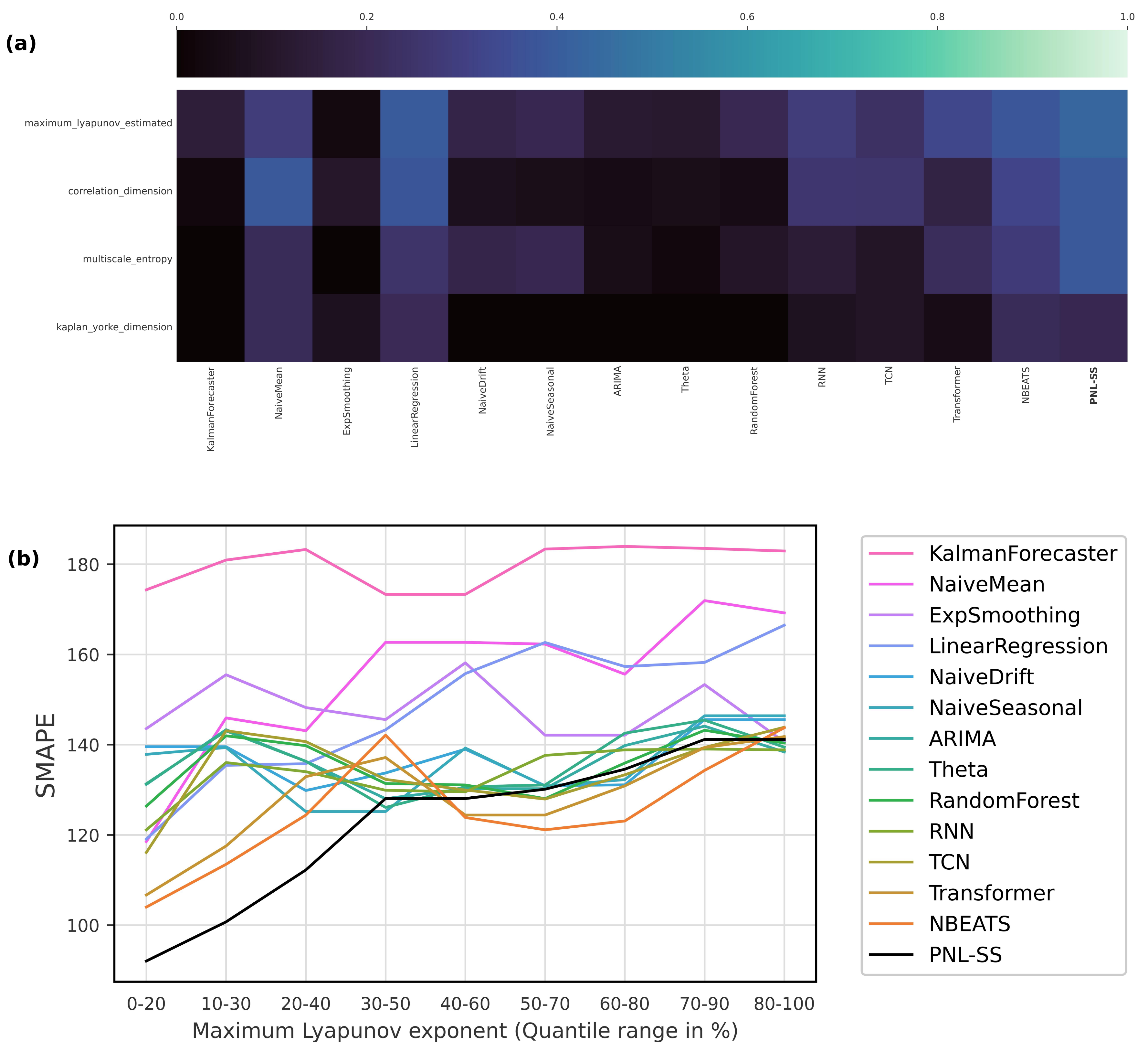}
\caption{Assessment of forecasting performance with respect to the properties of the chaotic systems. 
\textbf{(a)} The Spearman correlation between forecasting errors of each forecasting method and the underlying mathematical properties (Lyapunov exponent, correlation dimension, multiscale entropy, and Kaplan Yorke dimension) of the chaos systems. 
\textbf{(b)} The forecasting errors (symmetric mean absolute percent error, SMAPE) of different forecasting methods as a function of the maximum Lyapunov exponent of the dynamical systems represented in quantiles. The horizontal axis represents a sliding window that covers a 20 \% quantile of the distribution of the maximum Lyapunov exponent across all systems. Each point on the graph corresponds to the median performance across all dynamical systems in that quantile.}
\label{fig:combine2}
\end{figure}

We benchmarked the PNL-SS model by forecasting a univariate time series sampled from a large set of chaotic attractors with thirteen other major time series prediction methods. These models vary from classic methods such as ARIMA and Random Forest to deep-learning methods such as the long short-term memory (LSTM) \cite{salinas2020deepar}, Transformer \cite{vaswani2017attention,kim2021reversible}, and NBEATS \cite{oreshkin2019n}. Most deep-learning methods do not treat the time series as latent dynamics in their original formulation and may be limited to forecasting a one-dimensional signal and point predictions only. In contrast, PNL-SS predictions come with uncertainty estimates at the cost of increased computational complexity of learning and prediction. Nevertheless, here we compare our probabilistic model with these competing state-of-the-art architectures.

For this goal, we employed the scheme developed by William Gilpin \cite{gilpin2021chaos}, which standardized data sets of more than a hundred chaotic dynamical systems and benchmarked thirteen classical and deep-learning time series prediction methods based on the Darts environment for time series prediction \cite{herzen2022}. The dataset includes precomputed chaotic trajectories under different conditions. We selected the univariate train and test data sets that differ in the initial conditions under fine sampling granularity. See Appendix D and \cite{gilpin2021chaos, herzen2022} for further details on the dynamical systems and various other time series methods used for benchmarking. One advantage of using chaotic dynamics for benchmarking is that they are characterized by well-defined mathematical properties, such as the complexity of the trajectories, which enables us to elucidate the characteristics of each forecasting model.

To noisy observations of the 126 chaotic dynamics we fitted fourteen different time series prediction methods (including the PNL-SS method) whose hyperparameters are optimized by test datasets using the symmetric mean absolute percent error (SMAPE) as an error measure. To fit the PNL-SS model to one-dimensional time series data, we constructed delayed coordinates of $D_x$ dimension with the same number of latent space dimensions $D_x=D_y$. By doing so, we effectively introduced dependencies within the observations in $\*y_t$. We found that by providing this additional information to the model during training, the resulting forecasting performance improved. As additional hyperparameters, we considered the number of kernels and latent dimension from the combination of $L=[5, 10]$ and $D_x=D_y=[5, 30]$, and reported the models, that performed best on the validation set. See~\ref{appendix:forecasting details} for details.

Figure \ref{fig:combine1}a shows forecasts of AIZAWA (the first chaotic dynamics in the list of Gilpin's data sets) by exemplary methods, including methods based on deep learning. Note that, for the case of AIZAWA, the optimal number of the parameters selected in the LSTM-RNN, Transformer, and NBEATS were 8K, 548K, and 6.8M, respectively, whereas the PNL-SS model has only 420 parameters. Although the number of optimized parameters can differ for each dataset, the same numbers are typically chosen in other chaotic time series. Figure \ref{fig:combine1}b shows how the forecasting errors (SMAPE) of different models are related. The PNL-SS model (the bottom row) has a strong Spearman correlation with all the deep-learning models rather than the classical methods such as Exponential Smoothing or ARIMA, which means that it captures the underlying complexity in the time series in a similar way as a deep-learning model does, despite being simpler in design.

Figure \ref{fig:violin_high_low} shows how the forecast errors (SMAPE) are distributed for 126 dynamical systems and 14 forecasting models under high or low observation noise. We see that the PNL-SS model is one of the best models, along with the transformer and NBEATS model. For high noise levels (Figure~\ref{fig:violin_high_low}a), the PNL-SS model performs better than every other method. The plots shown are sorted by mean values of the SMAPE, but sorting by median values also made the PNL-SS take first place. To see the characteristics of the forecasting methods, we examined the forecasting error of each method with respect to several standard mathematical properties that measure the degree of chaotic behavior of a system (see \cite{gilpin2021chaos}). Figure \ref{fig:combine2}a shows the Spearman correlation between the forecasting errors and complexities of the trajectories measured by the Lyapunov exponent, correlation dimension, multi-scale entropy, and Kaplan-Yorke dimension, computed across all methods and all dynamical systems at high noise levels. Stronger positive correlations found for the PNL-SS model with these measures confirm that the performance of the model faithfully represented the complexities of the system than the other methods. Figure \ref{fig:combine2}b shows the relation between the Lyapunov exponent of the system and the forecasting error. It shows that the PNL-SS performs better for the systems with low Lyapunov exponents while the method is competitive with deep-learning methods for forecasting complex trajectories (i.e., high Lyapunov exponents), making it the best method on average for noisy conditions (Figure~\ref{fig:violin_high_low}a). 

\begin{figure}[tp]
\centering
\includegraphics[width=\linewidth]{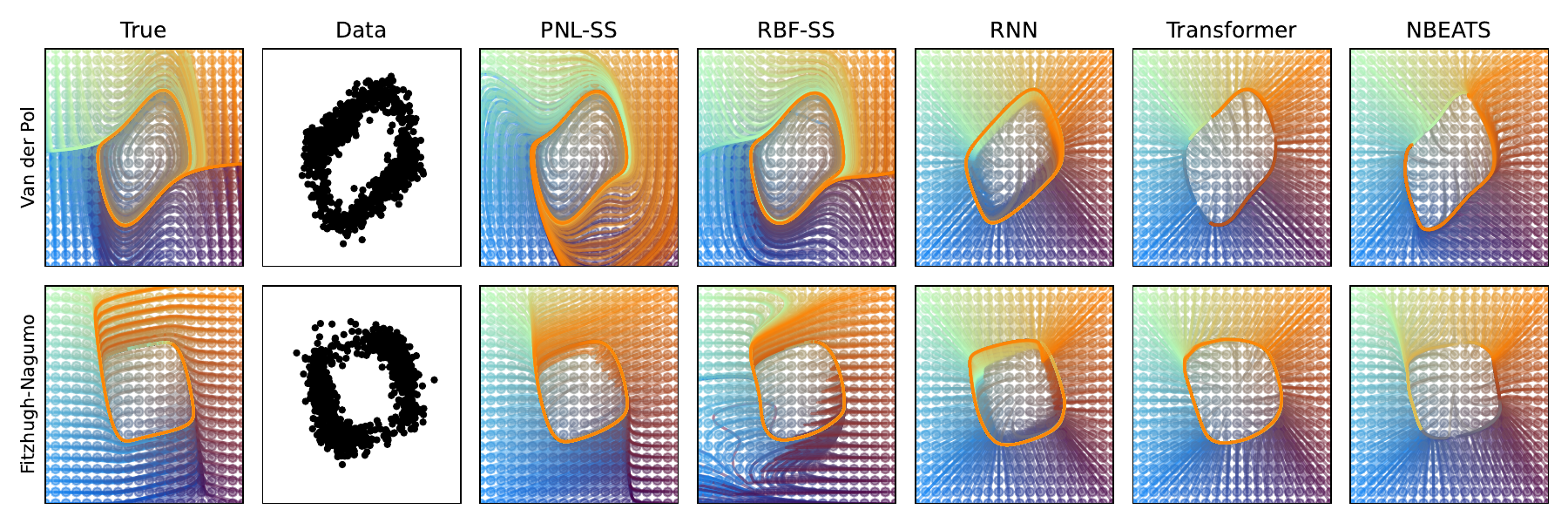}
\caption{Comparison of dynamical system recovery. From left to right, the underlying dynamical system, the noisy training data, and the reconstructed dynamics by PNL-SS, RBF-SS, RNN, Transformer, and NBEATS. The color of trajectories corresponds to its initial condition (colored dots). The top row is for the Van der Pol oscillator, and the bottom row is for the FitzHugh-Nagumo model. For details of the reconstruction method, see \ref{appendix:dynamical system comparison}.}
\label{fig:dyn_sys_comparison}
\end{figure}

Finally, we compared the models' ability to reconstruct the dynamical systems from noisy observations. Two simple two-dimensional systems, namely the Van der Pol oscillator and a FitzHugh-Nagumo model, were learned by the PNL-SS, RBF-SS, RNN, Transformer, and NBEATS model. Note that, in contrast to the previous prediction analysis, we limited the input of each model to the most recent time point for prediction. This potentially deteriorates the predictive performance (particularly for sequence-to-sequence models like transformer and NBEATS) but makes the dynamics in the phase space interpretable. After the model is trained, we predicted the dynamics for several initial conditions regularly spaced over the phase space (Figure~\ref{fig:dyn_sys_comparison}). Hence, we can assess how well the models learned the dynamics outside the regimes of the observed trajectories. For more details, see~\ref{appendix:dynamical system comparison}. 

The results demonstrate that all methods possess some ability to reconstruct the limit cycle. However, the predictions in regimes where no data were observed varied significantly. The PNL-SS and RBF-SS models recover the dynamics of the limit cycle and reasonably extrapolate the vector fields to the previously unseen regime. Figure~\ref{fig:vectorfield constructions}c provides visual intuition about how the linear and non-linear parts of the PNL-SS model (Eq.~\eqref{eq:state equation}) contribute to the vector field reconstruction. In contrast, the RNN, Transformers, and NBEATs fail to do so because, when starting from a point away from the limit cycle, their trajectories immediately jump to the regime of the training data. Thus, the deep-learning methods fail to extrapolate dynamics far away from the data.

\section{Discussion and summary}

The modeling of the nonlinear state function in our study (Eqs.~\eqref{eq:gauss_kernel} and \eqref{eq:scaler_input_to_nonlinear_func}) is in contrast to the popular choice of the Gaussian RBF and the square exponential (SE) function of the form, $\exp(-(\*x - \*c)^\top \bLambda (\*x - \*c))$ with $\bLambda$ being a scaled identity and full-rank matrix respectively. Our projection method can be considered as a restricted version of the SE function, where $\bLambda$ is a rank-one matrix, therefore reduces the distance along the one-dimensional vector, $\*w_{l}$,  to form a ridge function (Figure~\ref{fig:schamtic_illustration}\textbf{b}). We confirmed both theoretically and empirically this choice was computationally advantageous when performing inference and learning.

In addition to the computational efficiency, the choice of the rank-one matrix in the kernel describing the network of interactions is in agreement with the low-rank structure of the dynamics ubiquitously found in complex systems \cite{thibeault2024low}. Note that the Gaussian ridge function we employed is concentrated on the projected line defined by $\*w_l$ in the phase space but is non-concentrated in other directions (Figure~\ref{fig:schamtic_illustration}\textbf{c}), allowing generalization in broader regimes. However, we note that our model is a particular case of the models using the low-rank ridge functions, where we share the same kernels across the latent dimension. To explain this, consider a more general formulation in which the $i$th element of the nonlinear function is written as a linear combination of a bias term, a linear function, and kernel functions specific to each latent dimension:
\begin{align}
    f_i(\*x) = \*A_{{\rm lin},i} \bphi_{{\rm lin},i}(\*x) + \*A_{{\rm nl},i} \bphi_{{\rm nl},i}(\*x) + b_i.
    \label{eq:nonlinear_transtion_function}
\end{align}
Here the kernel vector $\bphi_{{\rm nl}, i}(\*x) = [\phi(h_{i,1}(\*x))),...,\phi(h_{i,L}(\*x))]^\top$ specific to the dimension is composed of
\begin{align}
    h_{i,l}(\*x) = \*w_{i,l}^\top \*x - {\tilde w_{i,l}}.
\end{align}
If $\*w_{i,l}$ is common for all kernels (i.e., $\*w_{i,l}=\*w_i$ for $l=1,\ldots,L$) and $\*A_{{\rm lin},i} = \mathbf{0}$ or $\*A_{{\rm lin},i} = \*w_i$, each nonlinear function $f_i(\*x)$ is a function of a scalar value $\*w_{i}^\top \*x$. Apart from the additive noise, this state model describes a one-layer continuous-valued recurrent neural network dynamics with synaptic weights $\*w_{i}$ and a nonlinear activation function $f_i(\*w_{i}\*x)$ ($i=1,\ldots,D_{\*x}$) \cite{valpola2002unsupervised}, where the function is determined by a non-parametric way rather than given as a parametric one such as a logistic/tanh or other types of sigmoid functions. This simplified form is still a generalization of the most kernel or GP-based state-space approach, where the same set of kernels is used for all latent states while only their coefficients in Eq.~\eqref{eq:nonlinear_transtion_function} are individually optimized \cite{zhao2020variational}. 

In this article, we explored the cases where the projection weights $\*w_{i,l}$ are different for each kernel while shared across latent variables: $\*w_{i,l}=\*w_l$ for all $i=1,\ldots,D_\*x$. This function can still be seen as a two-layer neural network if we consider each kernel an activation function of a neuron \cite{cybenko1989approximation,candes1999harmonic}. In this case, however, the nonlinear function  $f_i(\*x)$ for each latent dimension contains a greater number of the weight parameters than the case above, allowing us to capture a more detailed nonlinear transition of each latent variable, given that the number of kernels is the same. This indicates that the model potentially captures the nonlinear dynamics with fewer latent dimensions.

In summary, we provided a fast algorithm for inference and learning of nonlinear dynamics in time series data. One of the novel aspects of the proposed model is the use of kernel functions defined on projected lines to infer the nonlinear state dynamics. Our choice of the projected lines for each kernel while sharing them across latent dimensions makes the state model more expressive in capturing the dynamics with a fewer dimension while keeping the complexity of the model reasonably low. We showcased the model's ability to capture nonlinear dynamics and make accurate predictions using a prototypical model of noisy nonlinear oscillators and demonstrated its merit in computational efficiency and goodness-of-fit over the conventional RBF-based method. Finally, our exhaustive survey on forecasting 126 noisy, chaotic time series data revealed that our method outperforms deep-learning methods given the small samples and high noise, speaking to the alternative approach for forecasting time series. 

\section*{Code availability}
The python code for the projected nonlinear state-space model is available in the Github repository, 
\\ \noindent
\href{https://github.com/christiando/timeseries_models}{https://github.com/christiando/timeseries\_models}. 
\\The code uses the toolbox for operating a Gaussian distribution available in the Github repository, 
\\ \noindent
\href{https://github.com/christiando/gaussian-toolbox}{https://github.com/christiando/gaussian-toolbox}. 

\noindent
All code and data for figure generation are available for verification at
\\ \noindent
\href{https://github.com/christiando/pnlss}{https://github.com/christiando/pnlss}. 

\section*{Declaration of competing interest}
The authors declare that they have no known competing financial interests or personal relationships that could have appeared to influence the work reported in this paper.

\section*{Acknowledgments}
CD is supported by a Swiss Data Science Center project grant (C19-06). HS was supported by New Energy and Industrial Technology Development Organization (NEDO), Japan and JSPS KAKENHI Grant Number JP 20K11709, 21H05246.

\newpage
\appendix

\setcounter{figure}{0}
\setcounter{equation}{0}
\setcounter{section}{0}
\renewcommand{\thesection}{Appendix \Alph{section}}
\renewcommand{\theequation}{\Alph{section}.\arabic{equation}}

\section{Solving Gaussian integrals}\label{sec:gaussian integrals}
\label{appendix:gaussian_integrals}
In this section, we give the explicit solutions for expectations over functions involving the non-linear kernels $\phi_{{\rm nl},l}(\*x)$ with respect to a generic Gaussian density $p(\*x)$ with mean $\^\mu$ and covariance $\^\Sigma$.

First, the expectation is given as
\begin{equation}
    \EE{p(\*x)}{\phi_{{\rm nl},l}(\*x)} = \frac{\sqrt{|\tilde{\^\Sigma}_l|}\exp\left(\frac{1}{2}\tilde{\bmu}_{l}^\top \tilde{\^\Sigma}_l^{-1}\tilde{\bmu}_{l}\right)}{\sqrt{|\^\Sigma|}\exp\left(\frac{\tilde w_{l}^2}{2} + \frac{1}{2}\bmu^\top \bSigma^{-1}\bmu\right)}
\end{equation}
with  $\tilde{\^\Sigma}_{l} = \left[\^\Sigma^{-1} + \*w_{l}\*w_{l}^\top\right]^{-1}$ and $\tilde{\bmu}_{l} = \tilde{\^\Sigma}_{l}  \left[\tilde{w}_l\*w_{l} + \bSigma^{-1}\bmu\right]$. The second integral we encounter is of the following form
\begin{equation}
        \EE{p(\*x)}{\*x\phi_{{\rm nl},l}(\*x)} = \tilde{\bmu}_l\frac{\sqrt{|\tilde{\^\Sigma}_l|}\exp\left(\frac{1}{2}\tilde{\bmu}_{l}^\top \tilde{\^\Sigma}_l^{-1}\tilde{\bmu}_{l}\right)}{\sqrt{|\^\Sigma|}\exp\left(\frac{\tilde w_{l}^2}{2} + \frac{1}{2}\bmu^\top \bSigma^{-1}\bmu\right)}.
\end{equation}
The last integral we need to solve is over the product of two kernels
\begin{equation}
    \EE{p(\*x)}{\phi_{{\rm nl},l}(\*x)\phi_{{\rm nl},m}(\*x)} = \frac{\sqrt{|\tilde{\^\Sigma}_{lm}|}\exp\left(\frac{1}{2}\tilde{\bmu}_{lm}^\top \tilde{\^\Sigma}_{lm}^{-1}\tilde{\bmu}_{lm}\right)}{\sqrt{|\^\Sigma|}\exp\left(\frac{\tilde w_{l}^2 + \tilde w_{m}^2}{2} + \frac{1}{2}\bmu^\top \bSigma^{-1}\bmu\right)}
\end{equation}
with $\tilde{\^\Sigma}_{lm}^{-1} = \left[\*w_l\*w_l^\top + \*w_m\*w_m^\top + \^\Sigma^{-1}\right]$ and  $\tilde{\bmu}_{lm} = \tilde{\^\Sigma}_{lm} \left[\tilde w_{l}\*w_l + \tilde{w}_{m}\*w_m + \^\Sigma^{-1}\bmu\right]$. With these integrals, we can compute all the moments we need for learning the proposed algorithm.

\section{Details on filtering and smoothing procedure}
\label{appendix:filtering_smoothing}
In this section, we provide the mean and covariances for prediction, filtering and smoothing densities that we obtain by moment matching. The moments for the prediction density are straightforward to calculate with~\eqref{eq:prediction mean}--\eqref{eq:prediction covariance} and the results in Section~\ref{sec:gaussian integrals}. Since we use a linear observation model, we get the standard Kalman filter result for mean and covariance of the filtering density, i.e.,
\begin{align}
    \^\Sigma^{\rm f}_t = & \^\Sigma_t^{\rm p} - \^\Sigma^{\rm p}_t\*C^{\top}\^\Sigma_y^{-1}\*C\^\Sigma^{\rm p}_t,\\
    \bmu^{\rm f}_t = & \bmu^{\rm p}_t + \^\Sigma^{\rm p}_t\*C^{\top}\^\Sigma_y^{-1}(\*y_t - \*C\bmu^{\rm p}_t - \*d).
\end{align}
Once we iterated forward until $t=T$, we obtained the last {\it smoothing density} $p(\*x_T\vert \*y_{1:T})$, which is equal to the filtering density. To get the smoothing density for all the previous time-steps, we perform the following backward iteration. Assuming we have $p(\*x_{t+1}\vert \*y_{1:T})$, we want $p(\*x_t\vert \*y_{1:T})$.
To derive this density, we first need to calculate the lag-one density
\begin{equation}
    p(\*x_{t}\vert \*x_{t+1}, \*y_{1:t}) = \frac{p(\*x_{t+1}\vert \*x_{t})p(\*x_{t}\vert \*y_{1:t})}{p(\*x_{t+1}\vert \*y_{1:t})}.
\end{equation}
Because the numerator involves the non-linear state density, we again resort to moment-matching and get
\begin{equation}
    p(\*x_{t+1}, \*x_{t}\vert \*y_{1:t}) \approx {\cal N}\left(\begin{pmatrix}\bmu_{t+1}^{\rm p} \\ \bmu_{t}^{\rm f}\end{pmatrix}, \begin{pmatrix}\^\Sigma_{t+1}^{\rm p} & \*M_{t+1,t} \\ \*M_{t+1,t}^\top & \^\Sigma_{t}^{\rm f}\end{pmatrix}\right).
\end{equation}
For the covariances we get
\begin{equation}
    \*M_{t+1,t} = {\rm Cov}(\*x_{t+1},\*x_t) = \EE{p(\*x_{t}\vert \*y_{1:t})}{(\*A\phi(\*x_t) + \*b)\*x_t^{\top}} - \bmu_{t+1}^{\rm p}(\bmu_{t}^{\rm f})^{\top},
\end{equation}
which involves only integrals solved in \ref{sec:gaussian integrals}. With this we approximate the lag-one density by 
\begin{align}
    p(\*x_{t}\vert \*x_{t+1}, \*y_{1:t})\approx &{\cal N}(\bmu_{t}^{\rm f} + \*M_{t+1,t}^{\top}(\^\Sigma_{t+1}^{\rm p})^{-1}(\*x_{t+1} - \bmu_{t+1}^{\rm p}), \^\Sigma_{t}^{\rm f} - \*M_{t+1,t}^{\top}(\^\Sigma_{t+1}^{\rm p})^{-1}\*M_{t+1,t})\\
    = & {\cal N}(\*M_{t\vert t+1}^{\rm l}\*x_{t+1} + \*b^{\rm l}_{t\vert t+1}, \^\Sigma^{\rm l}_{t\vert t+1}),
\end{align}
with 
\begin{align}
\*M_{t\vert t+1}^{\rm l} = & \*M_{t+1,t}^{\top}(\^\Sigma_{t+1}^{\rm p})^{-1},\\
\*b^{\rm l}_{t\vert t+1} = & \bmu_{t}^{\rm f} - \*M_{t\vert t+1}^{\rm l} \bmu_{t+1}^{\rm p}.
\end{align}
With this, we get the approximation
\begin{equation}
    p(\*x_{t}, \*x_{t+1} \vert \*y_{1:T}) = p(\*x_{t}\vert \*x_{t+1}, \*y_{1:t})p(\*x_{t+1}\vert \*y_{1:T}) \approx {\cal N}(\bmu_{t,t+1}^{\rm s}, \^\Sigma_{t,t+1}^{\rm s}).
\end{equation}
From this, we can also get straightforwardly the marginal $p(\*x_{t}\vert \*y_{1:T})\approx {\cal N}(\bmu_{t}^{\rm s}, \^\Sigma_{t}^{\rm s})$.

\section{Details on calculating the Q-function}
\label{appendix:q_function}
For optimising the Q-function in Eq.~\eqref{eq:qfunction}, we again need to solve some Gaussian integrals. Since the initial density and the observation model are Gaussian, the first and second terms are straightforward to calculate. The last term involves the non-linear state model~\eqref{eq:state equation}, and hence we get for time $t$
\begin{equation}
\begin{split}
    & \EE{\vartheta}{\ln p(\*x_t\vert\*x_{t-1},\vartheta^\prime)} = \\
    & \EE{\vartheta}{-\frac{1}{2}\left(\*x_{t} - \*A\*\bphi(\*x_{t-1}) - \*b\right)^\top\^\Sigma_x^{-1}\left(\*x_{t} - \*A\*\bphi(\*x_{t-1}) - \*b\right)} - \frac{1}{2}\ln\vert2\pi\^\Sigma_x\vert.
\end{split}
\end{equation}
Again, we see that this factorizes into terms, that can be solved by integrals discussed in \ref{sec:gaussian integrals}. Also, it is straightforward to obtain the optimal $\*A,\*b$ and $\^\Sigma_x$ from this, by setting the gradients to zero. For the kernel parameters $\{\*w_l,\tilde{w}_l\}_{l=1}^L$, we find them numerically by the quasi-Newton method (L-BFGS-B) and automatic differentiation.

\section{Details on forecasting chaotic dynamics}
\label{appendix:forecasting details}
The benchmark of the PNL-SS model with other methods is based on the framework and code developed by William Gilpin \cite{gilpin2021chaos}. This framework provides standardized data sets of over a hundred chaotic dynamical systems and codes for benchmarking classical and deep-learning time series prediction methods. The benchmarking is built on the Darts environment, a Python library for time series analysis \cite{herzen2022}. We implemented our PNL-SS model as one of the forecasting methods callable by Darts. 

While we followed the procedures and parameter settings described in \cite{gilpin2021chaos}, we briefly summarize them below. Each one-dimensional chaotic time series data is composed of $1200$ time points, of which the first $1000$ points are used for training the model parameters, and the last $200$ points are used for testing the prediction. We added white Gaussian noise with zero mean and a standard deviation of $0.8$ for high noise conditions and $0.2$ for low noise conditions to the training data. The prediction error compared to the original data was computed using symmetric mean absolute percentage error (SMAPE) defined as $200 \langle {\left| z_t - \hat{z}_t \right|/(\left| z_t \right| + \left| \hat{z}_t \right|)} \rangle$, where $z_t$ is the test data and $\hat z_t$ is the forecasting values and $\langle \cdot \rangle$ denotes time average within a forecasting period (i.e., $t=1001,\ldots,1200$). The hyperparameters of the models, such as input and output chunk length or the number of network layers for deep-learning models, were optimized by the grid search function of the Darts by applying the same procedure above to data sets of the chaotic dynamics obtained separately from the evaluation procedure above. The ranges of the hyperparameters to be searched are unchanged from \cite{gilpin2021chaos}. This hyperparameter search was performed independently for each chaotic system under high and low noise conditions. In the analysis, we removed the chaos systems of GenesioTesi, Hadley, MacArthur, SprottD, and StickSlipOscillator due to errors encountered during ARIMA fitting. 

For the PNL-SS model, we embedded the time series data into the $D_y$-dimensional delayed coordinates and used it to train the model. A vector $\*y_t$ of the embedded trajectories at time point $t$ is composed of $D_y$ elements, each of which is a value shifted $200 / D_y$ points in time, while the last element is the original value at $t$. We used the same number of dimensions as the latent dimension ($D_x = D_y$). The last element in the vector was used for the prediction. For each chaos trajectory,  the embedding dimension $D_y$ and the number of kernels $L$ are hyperparameters, that we determined by grid search from the combination of $D_y=5$ or $10$ and $L=5$ or $30$. 

\section{Details on 2D dynamical system comparison}
\label{appendix:dynamical system comparison}
For creating Figure~\ref{fig:dyn_sys_comparison} we simulate data from two 2D systems. The first is the Van der Pol oscillator (Eqs.~\eqref{eq:VanderPol1}, \eqref{eq:VanderPol2}) with $\gamma=1$. 
The second is a FitzHugh-Nagumo system, with the governing equations
\begin{align}
    \dot{x}_1 & = 4 (x_1 - x_1^3 / 3 - x_2 + 0.7), \\
    \dot{x}_2 & = 4 (x_1 + 0.8 - x_2) / 12.5.
\end{align}
Both systems exhibit oscillatory behavior. The second system can be seen as an example of a ``stiff'' system, where $\dot{x}_1\gg\dot{x}_2$ or $\dot{x}_2\gg\dot{x}_1$, depending on the state of the system. These types of systems can be problematic for data-driven learning of dynamics~\cite{holt2022neural}. For both systems, we created a dataset, where we run the dynamics with initial condition $\*x(0)=[0,2]^\top$ by solving the equations numerically for the time points $t=0, 0.1, 0.2, \cdots, 119.8, 119.9$ resulting in $1200$ time steps. Then, we standardized each dimension and added Gaussian noise with $0.2$ standard deviation. This corresponds to the low noise condition in Figure~\ref{fig:violin_high_low}. The first 1000 data points of each dataset are then used for training. The last $200$ points are used only to evaluate the early-stopping criterion for the deep-learning methods.

In contrast to Figure~E.\ref{fig:violin_high_low}, we restrict all methods' input to the most recent time point of the data, since otherwise it was difficult to extract an interpretable phase space. The latent space of the Linear, the RBF-SS, and the PNL-SS model was set to $D_x=2$. For the latter two methods, we set $L=5$. To have an RNN model that is as close as possible to the PNL-SS, it is fit with 1 layer with 5 hidden units (similar to the number of kernels in PNL-SS). For the NBEATS and Transformer model, we use the default settings of the Darts library. All models are trained until convergence or until they reach 200 epochs. Figure~\ref{fig:vectorfield constructions} illustrates how each of the learned kernels of the PNL-SS model contributes to the reconstruction of the vector field of the Van der Pol oscillator.

\renewcommand{\thefigure}{E.\arabic{figure}}

\begin{figure}
    \centering
    \includegraphics[width=\linewidth]{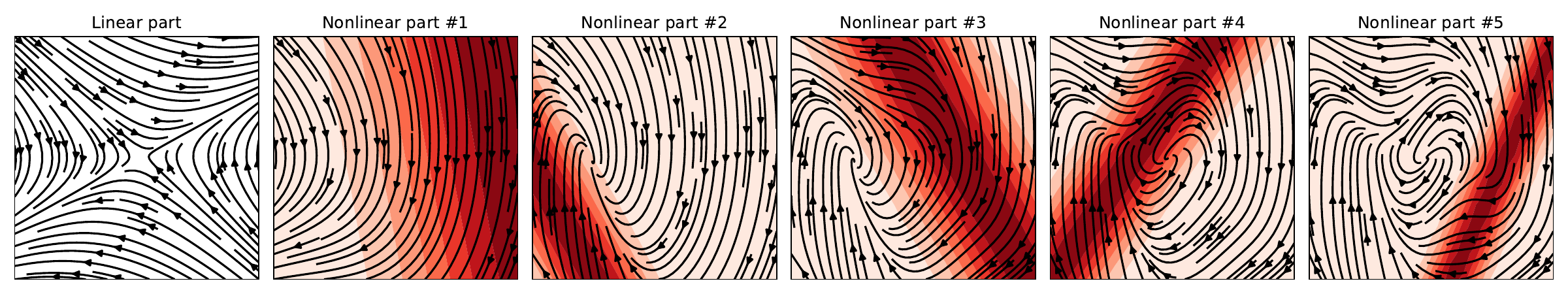}
    \caption{PNL-SS vector field reconstruction. From left to right: The streamlines in each panel depict the reconstruction of the vector field by the sum of the linear function and up to the $l^{\rm th}$ non-linear kernel functions. The order of the kernels is according to the norm of their weight vectors $\Vert \*A_{nl,l} \Vert$ ($l=1,\ldots,L$). The last panel shows the fully learned vector field. The red contour in each panel indicates the newly added kernel function $\phi_{nl,l}(\*x)$. 
    }
    \label{fig:vectorfield constructions}
\end{figure}

\newpage
\bibliographystyle{ieeetr}

\end{document}